\documentclass[onefignum,onetabnum]{siamart171218}

\usepackage{epsfig}
\usepackage{graphicx}
\usepackage{amsfonts}
\usepackage{amssymb}
\usepackage{cancel}
\usepackage{xcolor}
\usepackage{multirow}
\usepackage{subcaption}
\usepackage{lipsum}
\usepackage{epstopdf}
\usepackage{algorithmic}
\usepackage{soul}
\usepackage{xcolor}
\ifpdf
  \DeclareGraphicsExtensions{.eps,.pdf,.png,.jpg}
\else
  \DeclareGraphicsExtensions{.eps}
\fi

\definecolor{red_new}{RGB}{179,36,0}
\definecolor{green_new}{RGB}{0,128,43}

\newsiamremark{remark}{Remark}
\newsiamremark{hypothesis}{Hypothesis}
\crefname{hypothesis}{Hypothesis}{Hypotheses}
\newsiamthm{claim}{Claim}

\headers{Fixing match-fixing: Optimal schedules to promote competitiveness}{M. Chater, L. Arrondel, J-P. Gayant and J-F. Laslier}

\title{Fixing match-fixing: Optimal schedules to promote competitiveness\thanks{Submitted to the editors DATE.}}

\author{Mario Chater \thanks{Paris School of Economics, France
  (\email{mario\_chater@hotmail.com}), corresponding author.}
 \and Luc Arrondel \thanks{Paris School of Economics, France
  (\email{luc.arrondel@gmail.com}).}
\and Jean-Pascal Gayant \thanks{GAINS, Le Mans Université, and CREM, Université Rennes 1, France (\email{jean-pascal.gayant@univ-lemans.fr})}
\and Jean-François Laslier \thanks{Paris School of Economics, France (\email{jean-francois.laslier@ens.fr})}}

\usepackage{amsopn}

\ifpdf
\hypersetup{
  pdftitle={Fixing Match-Fixing},
  pdfauthor={M. Chater, L. Arrondel, J-P. Gayant and J-F. Laslier}
}
\fi

\begin{document}

\maketitle

\begin{abstract}
In the last round of the FIFA World Cup group stage, games for which the outcome does not affect the selection of the qualified teams are played with little enthusiasm. Furthermore, a team that has already qualified may take into account other factors, such as the opponents it will face in the next stage of the competition so that, depending on the results in the other groups and the scheduling of the next stage, winning the game may not be in its best interest. Even more critically, there may be situations in which a simple draw will qualify both teams for the next stage of the competition. Any situation in which the two opposing teams do not play competitively is detrimental to the sport, and, above all, can lead to collusion and match-fixing opportunities. We here develop a relatively general method of evaluating competitiveness and apply it to the current format of the World Cup group stage. We then propose changes to the current format in order to increase the stakes in the last round of games of the group stage, making games more exciting to watch and, at the same time, reducing any collusion opportunities. We appeal to the same method to evaluate a ``groups of 3" format which will be introduced in the 2026 World Cup edition as well as a format similar to the one of the current Euro UEFA Cup.

\end{abstract}



\newpage
\section{Introduction}
\textbf{} \\
The soccer FIFA World Cup is a global sporting event that attracts one of the highest audiences: according to the organizing entity, FIFA, over one billion television viewers watched the final between France and Croatia\footnote{The official report can be found at \textit{https://resources.fifa.com/image/upload/the-2018-fifa-world-cuptm-in-numbers.pdf?cloudid=veij99mubas9idvf47rl}.}on $15^{th}$ July 2018.
To further improve the worldwide World Cup audience, FIFA is currently trying to include more countries in the final stages of the tournament and increase the attractiveness of all of the games played during this one-month competition.

In its current format, the World Cup consists of 32 qualified teams (via continental qualifying tournaments) that are distributed into 8 groups of 4 teams each. At the group stage, teams in the same group play against each other once (for a total of 6 matches per group) with the group ranking based on the current football points system: 3 points for a win, 1 for a draw and 0 for a loss. The first two teams in each group qualify for the knockout stage.

This schedule can lead to games being played in an unethical and unattractive way, with a good example being the infamous match between Austria and West Germany (Germany being at that time still divided) during the 1982 FIFA World Cup, in Gijon (Spain). Sometimes called the ``disgrace of Gijon", the game is known in Germany and Austria as the ``Gijon non-aggression pact". West Germany and Austria both played in group 2, with Algeria and Chile. To everyone's surprise, the German \textit{Mannschaft} stumbled against Algeria, losing its first match (1-2) (the first time that a European team had lost to an African team in a World Cup), and Austria beat Chile (1-0). In the second round of games, West Germany beat Chile (4-1) while Algeria lost to Austria (0-2). At this point, Chile was already eliminated. The last two games in Group 2 were thus decisive for Germany, Austria and Algeria. On June 24, Algeria beat Chile 3-2 and the standing of the group became as in Table \ref{Table_Gijon_1}\footnote{In the 1982 version of the World Cup, 2 points were awarded for a win, 1 for a draw, 0 for a loss. The adoption of the current football points system (3 points for a win) would have led to an identical situation and could have not eliminated the collusion opportunity.}. From then on, an arrangement became possible between the West Germans and Austrians. A simple calculation shows that, with a simple 1-0 victory (or even 2-0 victory) for the \textit{Mannschaft}\footnote{ In case of equality in the number of points between teams, the first tie-breaker was goal difference.}, both the Germans and Austrians would qualify while, oddly, either a large German victory, an Austrian victory or a draw would lead to Algeria qualifying. After 10 minutes of game, the Germans scored, following which both teams almost stopped for the remaining, long, 80 minutes, under the booing of infuriated Spanish spectators. After the scandal of Gijon (and similar games), the last two games in each group are now played simultaneously, but this has not completely eliminated collusion opportunities. Be it illegal match-fixing or tacit collusion, such incident is very detrimental to the game. At the other end of the spectrum, we can think of a situation where the two opposing teams played honestly (competitively) and the result of the game ended up benefiting both. We can clearly expect that both teams will be criticized and their reputation will be tarnished after this game (in the same way that a tacit collusion cannot be proved for sure, there is no way the teams can prove they played competitively). Either way, FIFA should reduce to the extent possible these situations of ``potential" match-fixing since, even if match-fixing did not occur, it is the existence of such a possibility in the minds of spectators and players that is detrimental to sportsmanship.\\
All things equal, we deem that a tournament schedule is more ``fair" if the probability of reaching situations that encourage match-fixing is lower. To be specific, by match-fixing, we mean the action of predetermining the outcome of a game in a way that one team or both play intentionally below their actual potential in order to allow the other team to score/not to concede a goal. Tournament fairness is a broad concept and has been vastly discussed. Appleton \cite{Appleton}, Scarf et al. \cite{Scarf} and Lasek and Gagolewski \cite{Lasek} define a fair tournament as one in which the intrinsically best teams have the highest probability of winning. In other words, the tournament is designed such that it reveals the strongest teams. In addition to improving our own ``metric" of fairness, our proposed optimal tournament scheduling preserves if not improves fairness as defined in the above literature. Based on a classification method that we develop and present later in this article, a game is either \textit{competitive}, \textit{stake-less} or \textit{collusive}. We argue that match-fixing opportunities (when there is an incentive for a match to be fixed) occur rarely in stake-less games, always in collusive games and never in competitive games (as the preferred outcome for one team is incompatible with the one of the other team). By increasing the number of competitive games we are able to minimize the types of games which may lead to match-fixing. It should be noticed that the maximization of competitive games is at the same time a solution as well as a by-product of our initial objective: having more competitive games is, per se, beneficial to the sport since they are exciting to watch. The assessment/simulation method we develop allows us to pick an optimal game schedule but does not alter any other aspect of the tournament such as the number of teams in the group, games played, qualified teams, points system, etc. Consequently, given the final results of all games in the group, the final ranking will remain untouched regardless of the tournament schedule meaning that the fairness metric as defined in Appleton \cite{Appleton} is preserved. On the contrary, a schedule that increases the number of competitive games helps eliminate "psychological bias" (not covered by Appleton's metric). By imposing a sequence of games which forces the teams to give their best, it increases the chances of the tournament revealing the strongest.

\begin{table}[!htb]
    \begin{minipage}{.5\linewidth}
      \caption{Group ranking before the last game between West Germany and Austria (MP = matches played)}\label{Table_Gijon_1}
      \centering
        \begin{tabular}{|l|c|c|c|}
        \hline
        Team & Pts & Goal dif. & MP\\
        \hline
        Austria & 4 & +3 & 2\\
        Algeria & 4 & 0 & 3\\
        West Germany & 2 & +2& 2\\
        Chile & 0 & -5 & 3\\
        \hline   
        \end{tabular}
    \end{minipage}%
    \begin{minipage}{.5\linewidth}
      \caption{Final group ranking}\label{Table_Gijon_2}
      \centering
        \begin{tabular}{|l|c|c|}
        \hline
        Team & Pts & Goal dif.\\
        \hline
        West Germany & 4 & +3\\
        Austria & 4 & +2 \\
        Algeria & 4 & 0 \\
        Chile & 0 & -5\\
        \hline   
        \end{tabular}
    \end{minipage}%
\end{table}

FIFA defines the groups of the World Cup through a draw procedure that changes slightly from year to year depending on the origin of the qualified teams (FIFA tries to spread out teams from the same continent in an even manner across all of the groups). Nevertheless, the main structure of the draw is the following:
\begin{itemize}
\item Teams are divided into pots, each of which is supposed to contain teams with similar levels of performance: the 8 best teams in pot A, the second best 8 teams in pot B, and so on. Team performance is based on a ranking whose methodology has changed over time, and has been criticized by some football experts \cite{Cea_Duran}, \cite{Gasquez}. The country hosting the tournament (which qualifies automatically) is included in pot $A$ in order to maximize its chances of proceeding to the knockout stage of the tournament.
\item Groups are formed by picking one team from each pot so that all groups have a ``top-level" pot-$A$ team, a ``second-level" pot-$B$ team, a pot-$C$ team and a ``weaker" pot-$D$ team.
\item Finally, the schedule of the games, i.e. the order in which the teams play against each other is drawn randomly. As such, in some groups the last round of games will consist of ``pot $A$" vs ``pot $B$", and ``pot $C$" vs ``pot $D$", while in other groups the last matches consist of ``pot $A$" vs ``pot $C$", and ``pot $B$" vs ``pot $D$", or ``pot $A$" vs ``pot $D$" and ``pot $B$" vs ``pot $C$".
\end{itemize}

\medskip

Having in mind our target of eliminating situations of ``potential" match-fixing, we develop a method to evaluate the competitiveness of the last-round games, choose a model to simulate the group-stage outcomes and look for the optimal setting that maximizes our competitiveness metric (which is equivalent to  minimizing the ``potential" match-fixing events). We also apply this method to real World Cup data (starting from the 1998 competition, in which the new format was adopted) and conclude that games were sub-optimally scheduled. We find out that the points-attribution scheme does not affect the quality of games in that a victory produces more points than a draw, which produces more points than a loss. However, the order in which games are played, and specifically the schedule for the last round of games, is critical and can substantially improve the competitiveness of the last round if well-designed. The methodology we have developed for assessing competitiveness is pretty general and can be applied to any tournament structure based on rankings and tie-breaking rules\footnote{Such tournament structures can be found in many other sports such as basketball, rubgy, volleyball, handball and even in non-sporting activities.}.\\

Our work gives continuity to a growing litterature that looks at the competition design of sporting tournaments from a scientific point of view as presented by Kendall and Lenten \cite{Kendall}. Based on backward induction analysis, Krumer et al. \cite{Krumer1} showed that in round-robin tournaments among three or four symmetric contestants, there is a first-mover advantage driven by strategic effects arising from the subgame perfect equilibrium. This article is a foundation for an empirical analysis carried by Krumer and Lechner \cite{Krumer2} applied to different sport events including the FIFA World Cup. Csato \cite{Csato} demonstrates the incentive incompatible design of recent UEFA qualification tournament which includes a repechage procedure. In a study applied to Super Rugby but which can be adapted to football, Winchester \cite{Winchester} determines the optimal allocation of points that most appropriately rewards strong teams. Furthermore, Brams and Ismail \cite{Brams_Ismail}, Anbarci, Sun and Unver \cite{Anbarci} design new penalty shoot-out approaches which improve fairness at the knock-out stage of the World Cup. Related to the penalty shoot-out, Lenten, Libich and Stehlik \cite{Lenten} explore a better tie-breaker mechanism altogether. Concerning the design of the draw procedure, Guyon \cite{Guyon1} and Laliena and Lopez \cite{Laliena} analyze the fairness of the FIFA World Cup draw and propose new draw schemes which are more equitable and preserve the geographic constraints enforced by FIFA. Finally, Guyon \cite{Guyon2} tackles the match scheduling of group games of the 2026 FIFA World Cup (16 groups of 3) and points out that this format of competition makes the "disgrace of Gijon" possible again. He suggests that the strongest team in the group should be the one to play the first two games of the group in order to minimize the risk of collusion in the last round game. In the same spirit as in Guyon's study \cite{Guyon2}, our article analyzes the tournament structure of the group stage having in mind the final target of reducing collusion opportunities and increasing competitiveness. We develop a rather general theoretical framework to assess the competitive level of games and combine it with a model which simulates game scores to come up with an exact quantitative assessment of any group format. We fit our simulation model to historical FIFA World Cup data, and evaluate the current format of the competition as well as potential future formats (recovering the optimal scheduling suggestion of Guyon for the ``groups of 3" format). 

This article is organized as following: we first benchmark and calibrate different score-predicting models, based on the results in previous World Cups and team rankings. We next develop a classification method that allows us to quantify the attractiveness of the last round of games. Based on the chosen model and our original method, we then use Monte Carlo simulations to determine the key factors that affect the quality of the last round of games in the current World Cup format and propose a remedy to improve the competitiveness of the last round of the group stage. The last section applies our method to the new enlarged version of the World Cup. We here analyze the two ``well-known" options: 16 groups of 3 teams and the current "UEFA Euro"-format with groups of 4 groups with ``repechage" of the best third-ranked teams as well as 2 variants of an 8 groups of 5 alternative format.

\section{Group-stage model}
\textbf{} \\
In the past decades, many models have been developed to predict or simulate the outcome of football games. For example, Lee \cite{Lee} and Dyte and Clarke \cite{Dyte_Clarke1} treat the goals scored by each team as conditionally-independent Poisson variables whose parameters depend on team attributes and the match venue. Maher \cite{Maher} found that introducing a correlation between the number of goals scored via a bivariate Poisson distribution improved predictive power in data from the English League. Reep, Pollard and Benjamin \cite{RPB} construct a model based on the negative binomial distribution, while Karlis and Ntzoufras \cite{Karlis_Ntzoufras} use Skellam's distribution to model the difference in the number of goals (the margin of victory). In a following article, the latter develop a robust fitting method to account for abnormal large scores \cite{KN2}. Most of this previous work has looked at national championships, but only few have considered the FIFA World Cup. The 1998 World Cup is covered in \cite{Dyte_Clarke1}. Suzuki et al. \cite{Suzuki} use a Bayesian approach to predict the result of the 2006 World Cup, while Groll et al.  \cite{Groll} apply their model to the 2014 World Cup. \\

We here benchmark the different models that will be used to simulate the game outcomes. For the sake of simplicity, a team's strength is completely described by one single variable. Instead of using FIFA rankings, we choose the Elo index as a proxy for team performances\footnote{All historical and current Elo ratings, as well as the details on how they are calculated, can be found at \textit{https://www.eloratings.net}}. The advantage of this index is that it is more transparent\footnote{Elo ratings are continuous rather than ordinal, so that they allow for different rating ``gaps" between consecutively-ranked teams.} and is a more accurate reflection of a team's real level than the FIFA ranking \cite{Gasquez}. The calculation method has not changed over time, and it is thus better-suited for analysis over long time periods (we will here cover all the World Cups starting from that in 1998). Nevertheless, the Elo and FIFA point systems produce very similar country rankings.\\

Each team has an Elo index that belongs to the interval $[a,b]$, with $a$ being the lowest Elo rating while $b$ is the highest. As in the official draw procedure, we form groups of four teams with different Elo indices as follows: team $A$'s Elo is uniformly drawn from the interval $[b-\frac{b-a}{4},b]$, team $B$ from $[b-\frac{b-a}{2},b-\frac{b-a}{4}]$, team $C$ from $[a+\frac{(b-a)}{4},b-\frac{b-a}{2}]$ and team $D$ from $[a,a+\frac{b-a}{4}]$. In other words, pot $A$ is constituted of the best ranked teams while pot $D$ includes the lowest ranked teams. The bounds $a$ and $b$ are parameters in our model and will be calibrated in the next section. Intuitively, the greater the $b-a$ gap, the larger the performance gap between teams within the group. Based on the Elo indices of the teams in the  group, we simulate the outcomes of their matches.

\subsection{Simulating match outcomes}
\textbf{} \\
As we only consider a team's Elo index, we analyze the following relatively simple parametric models:
\begin{enumerate}
\item \textbf{Simple Poisson model}: Each time a team has the ball it can attack and score a goal. With $n$ attack opportunities and a probability $p$ of scoring per attack, the number of goals scored follows a binomial distribution $\mathcal{B}(n,p)$. On average, $\lambda=n\cdot p$ goals will be scored by the team per game. The binomial distribution limits the number of goals scored per game to the total amount of attacks $n$. If instead of considering discrete attacks, we look at ball possession and introduce the probability of scoring per unit of ball possession, $\lambda$, the number of goals scored follows a Poisson distribution with parameter $\lambda = n \cdot p$. This distribution is the limit case of the binomial distribution as $n\to \infty$ (every ball possession signifies an attack) and $p=\frac{\lambda}{n}\to 0$ (the strict probability becomes a probability density of scoring per unit of time possession). The probability that team $i$ scores $k$ goals against team $j$ is:
\begin{align}
P(goals=k)=\frac{\lambda^k\cdot e^{-\lambda}}{k!}
\end{align}
where $\lambda$ is:
\begin{align}
\lambda=\alpha\cdot \frac{r_i}{r_i+r_j}
\end{align}
Here $r_i$ is the Elo index of team $i$, $r_j$ the Elo index of its opponent $j$, and $\alpha$ a parameter of the model to be calibrated. The stronger the scoring team (the higher is $r_i$) and the weaker its opponent (the lower is $r_j$), the easier it will be for team $i$ to score goals. The parameter $\alpha$ reflects how prolific games are in terms of goals scored (a higher $\alpha$ produces games with higher scores). Consequently, the result $(k_i,k_j)$ of a game between team $i$ with Elo index $r_i$ and team $j$ with Elo index $r_j$ is distributed:
\begin{align}
(k_i,k_j)\sim (X,Y)
\end{align}
where X and Y are independent, and $X\sim \mathcal{P}(\alpha\cdot \frac{r_i}{r_i+r_j})$ and $Y\sim \mathcal{P}(\alpha\cdot \frac{r_j}{r_i+r_j})$. We can notice that such a model implies that each match yields $\alpha$ goals on average which are shared between the teams based on their ratings. Assuming that each match has the same average number of goals is not an unreasonable assumption in our context. First, we will only simulate outcomes for the first two rounds of the group stage during which the teams usually follow their ``baseline" tactic. Games in which a team will opt for an aggressive tactic (exposing himself to counter-attacks and leading to games with a high number of goals) or, on the contrary, to a very "defensive" tactic (leading to games with almost no goals) usually happen in the third round of the group stage where, under some circumstances, tie-breaking rules such as goal difference or the number of goals scored start driving the behavior of teams. Furthermore, the style of football may have evolved\footnote{Games used to have more goals in the earliest editions of the World cup, with averages above 4 goals per game.} slightly over the period covered by our sample (1998 to 2018) but not enough to have had a significant impact on the average number of goals scored per game (as per Table \ref{goals_per_game}).
      
\item \textbf{Bivariate Poisson model}: The bivariate Poisson model is very similar to that above. The only difference is that it accounts for correlations between the number of goals scored by the two teams. The underlying idea here is that if one team scores, the other will attempt to equalize and put more effort into scoring. This leads to open games with a greater number of goals on both sides. On the contrary, if neither team scores the game will remain ``closed" with few goals. The final score of the game $(k_i,k_j)$ is:
\begin{align}
k_i=X+Z \qquad k_j=Y+Z
\end{align}
where $X$, $Y$ and $Z$ are independent, and $X\sim \mathcal{P}(\alpha\cdot \frac{r_i}{r_i+r_j})$, $Y\sim \mathcal{P}(\alpha\cdot \frac{r_j}{r_i+r_j})$, $Z\sim \mathcal{P}(\beta)$. The correlation between $k_i$ and $k_j$ comes from the term $Z$, and the greater is $\beta$ the higher the correlation. This model has one more parameter than that above (namely $\beta$). Based on its specification, this model also assumes that each match yields $\alpha+2\beta$ goals on average. 

\item \textbf{Negative binomial model}: In this model, when one team scores a goal, it becomes more motivated and has a greater probability of scoring a second goal. The scoring model starts as a Poisson distribution, and each time a goal is scored the probability of scoring the next goal rises by a given constant. The probability distribution of goals can be calculated and has a negative binomial distribution. The probability that team $i$ scores $k_i$ goals against team $j$ is:
\begin{align}
P(k_i)= \begin{pmatrix}
k_i+r-1\\
k_i
\end{pmatrix} \cdot (1-\alpha\cdot \frac{r_i}{r_i+r_j})^r\cdot (\alpha\cdot \frac{r_i}{r_i+r_j})^{k_i}
\end{align}
where $r\in \mathbb{N}$ and $\alpha>0$ are two parameters to be calibrated.

\item \textbf{Ordered logistic regression (OLR)}: as per Hvattum and Arntzen \cite{Hvattum}, we fit an OLR model to our data using the Elo indices as regressors. As opposed to the previous models, OLR does not predict exact scores but assigns probabilities for the three ``compact'' outcomes of a football game: first team wins the game, draw, first team losses the game. Since this model is specifically designed to classify games into the three above categories, we expect it to outperform the previous three on the learning sample when the evaluation criteria is only sensitive to whether the outcome is a win, draw or loss. Consequently, the OLR model would set a benchmark to which we can compare the performances of the previous models (only for "coarse" criteria that disregard the exact final score of the game).      
\item \textbf{Naive uniform guess}: as its name suggests, each of the three ``compact" outcomes has the same probability of happening (1/3) regardless of the Elo indices of the competing teams. This ``model" goes into the same category as OLR since it cannot deal with exact scores and has only been introduced as a ``worst case" benchmark. 
\end{enumerate}

\begin{center}
\begin{table}
\begin{tabular}{|c|c|c|c|c|c|c|}
\hline
World Cup edition & 1998 & 2002 & 2006 & 2010 & 2014 & 2018\\
\hline
Average number of goals per game & 2.6 & 2.7 & 2.4 & 2.1 & 2.9 & 2.5 \\
 \hline
\end{tabular}
\caption{Average number of goals scored per game during the group stage of FIFA World Cups from 1998 to 2018}\label{goals_per_game}
\end{table}
\end{center}

\subsection{Rescaling the Elo distribution}
\textbf{} \\
The Elo indices usually fluctuate between 1500 and 2200 for the teams that qualify for the World Cup (Table \ref{Elo description} presents some descriptive statistics of the Elo index in our sample). As such, $\frac{r_i}{r_i+r_j}$ varies between $\frac{1500}{1500+2200}=0.4054$ and $\frac{2200}{1500+2200}=0.5946$. Consequently the ``raw" Elo indices will barely affect the number of goals scored by teams. We amplify the performance difference between teams via a linear transformation of the original Elo indices:
\begin{align}
r_i'=1+e^{gap}\cdot\frac{r_i-\min_j(r_j)}{\max_j(r_j)-\min_j(r_j)}
\end{align}
After this transformation, the weakest team will have an index of 1 and the strongest an index of $1+e^{gap}$, where $gap$ is a parameter to be calibrated\footnote{The introduction of $e^{gap}$ instead of a simple linear function of $gap$ has been chosen solely for practical purposes: ensuring that it is positive and to increase the sensitivity of our model to the $gap$ parameter (which can improve the convergence speed of the maximum-likelihood search algorithm).}. The higher is $gap$, the larger the performance gap between teams and the greater the impact of the Elo indices on a game's outcome.

\begin{table}
\begin{center}
\begin{tabular}{|c|c|c|c|c|}
\hline
Minimum & 1st quantile & Median & 2nd quantile & Maximum\\
\hline
 $1491$ & $1714$ & $1815$ & $1920$ & $2142$  \\
\hline
\end{tabular}
\caption{Descriptive statistics of the Elo indices of our sample.}\label{Elo description}
\end{center}
\end{table}

\subsection{Model selection and calibration}
\textbf{} \\
We now carry out maximum-likelihood estimation of the three exact models presented above as well as the OLR, in order to decide which will be used to carry out our group simulations. The data used to fit these models are the results of the first two rounds of the World Cup group stage from 1998 up to 2018. This covers 192 games: the last round of games is not included as factors other than team performance may play a role (teams may prefer to lose or draw in the last game, and finish second in the group in order to have easier knock-out games). Table \ref{model selection} shows the results of our estimations. In a second step, we have chosen different metrics to evaluate the performance of the models. For the metrics ``Log-likelihood", ``AIC" and ``BIC" we decided to restrict ourselves to the first three models and only show figures that can actually be compared: the first three models predict exact final scores (which is not the case for the last two models which can only predict wins-draws-losses). On the other hand, the ``Logloss" and ``Brier score" evaluate the performance of each model in predicting the win-draw-loss outcome of a game and allow for a broader comparison of the models. The two latter measures estimate a ``distance" between the predicted outcome   of a model and the actual result (a rigorous presentation of these loss functions can be found in \cite{Witten_Frank})  
\begin{center}
\begin{table}
\begin{tabular}{|c|c|c|c|c|c|c|}
\hline
\multicolumn{2}{|l|}{\textbf{Model}} & Simple Poisson & Bivariate Poisson & Negative Binomial & OLR & Uniform\\
\hline
\multirow{4}{*}{ \textbf{Optimal parameter values}}& $gap$ & $3.7581$ & $3.7582$ & $3.4997$ & \multirow{8}{*}{} & \multirow{8}{*}{}\\
 & $\alpha$ & $2.5156$ & $2.5156$ &  $0.1747$ & &\\
 & $\beta$ &-& $2.7518\cdot 10^{-10}$&- & &\\
 & $r$ &-&-& $13$ & & \\
 \cline{1-5}
 \multicolumn{2}{|l|}{\textbf{Number of parameters}} & $2$ & $3$ & $3$ & &\\
 \cline{1-5}
 \multicolumn{2}{|l|}{\textbf{Log-Likelihood}}& $-535.6698$ & $-535.6698$ & $-534.4337$ & &\\
 \cline{1-5}
 \multicolumn{2}{|l|}{\textbf{AIC}} & $1075.3$ & $1077.3$ & $1074.9^*$ & &\\
 \cline{1-5}
 \multicolumn{2}{|l|}{\textbf{BIC}} & $1081.9^*$ & $1087.1$ & $1084.6$ & &\\
 \hline
 \multicolumn{2}{|l|}{\textbf{Logloss}} & $0.9498$ & $0.9498$ & $0.9491$ & $0.9334$ & $1.0986$\\
 \hline
 \multicolumn{2}{|l|}{\textbf{Brier score}} & $0.1882$ & $0.1882$ & $0.1880$ & $0.1832$ & $0.2222$\\
 \hline
\end{tabular}
\caption{In-sample estimation results with different models for the  prediction of game scores}\label{model selection}
\end{table}
\end{center}

\begin{figure}
\centering
\begin{subfigure}{.5\textwidth}
  \centering
  \includegraphics[width=1\linewidth]{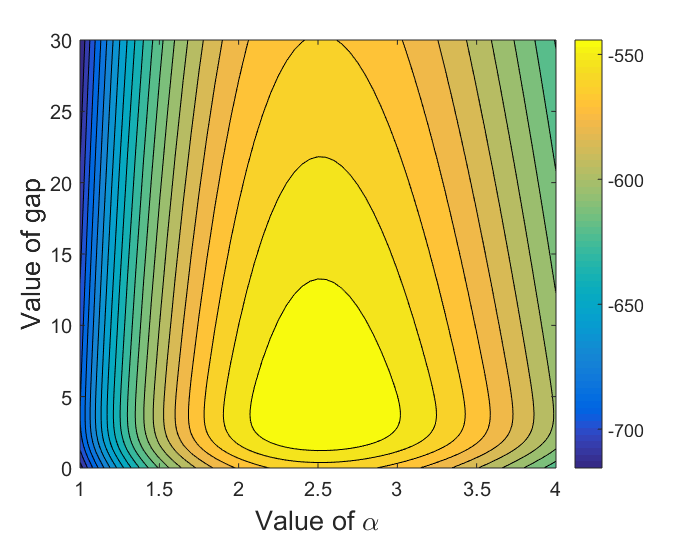}
  \caption{Poisson distribution}
\end{subfigure}%
\begin{subfigure}{.5\textwidth}
  \centering
  \includegraphics[width=1\linewidth]{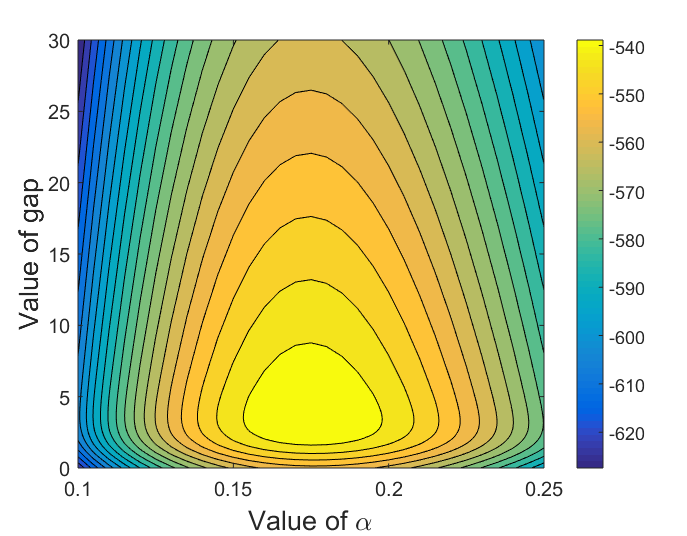}
  \caption{Negative-Binomial distribution}
\end{subfigure}
\caption{The Log-likelihood of different models as a function of their parameters}
\label{contourfs}
\end{figure}

First, since the optimal value of $\beta$ is $2.7518\cdot 10^{-10}$, we conclude that the introduction of a correlation term between the goals does not improve the accuracy of the simple Poisson model (there is no change in the log-likelihood between the simple and the bivariate Poisson models either). Furthermore, the value of $\alpha$ which maximizes the likelihood is equal to the average number of goals per game of our sample ($2.5156$ goals per game). This result serves as an additional check since $\alpha$ (resp. $\alpha+2\beta$) is equal to the average number of goals per game in the simple Poisson (resp. bivariate Poisson) model. In addition to that, we can notice that the optimal value of $gap$ is strictly positive, so that the Elo indices do have predictive power for our game outcomes (Figure \ref{contourfs} shows the likelihood in the simple Poisson and Negative-Binomial distributions as a function of the model parameters).\\

As for the logloss and Brier scores, we notice that our ``score-predicting" models are almost as efficient in predicting win-draw-loss outcomes for the given sample than the specifically-designed OLR model.
Among these ``exact" models, the simple Poisson model minimizes BIC since it has one less parameter than the other two while the negative binomial model minimizes AIC.\\
For the following part of the paper, we decide to adopt the simple Poisson model for its tractability and great performance: its parameters can be clearly interpreted and it offers identical performance to the other two exact-score models. In terms of predicting win-draw-losses, its performance is very close to the specifically-designed OLR. Last but not least, the Poisson model is picked instead of OLR because having an exact-score generating model sets the ground for research work requiring exact scores (including goal differences and other goals-linked tie-breaking rules) which is the case in the second part of this study.


\subsection{Additional performance checks for the Poisson model}
\begin{figure}
\centering
\includegraphics[scale=0.5]{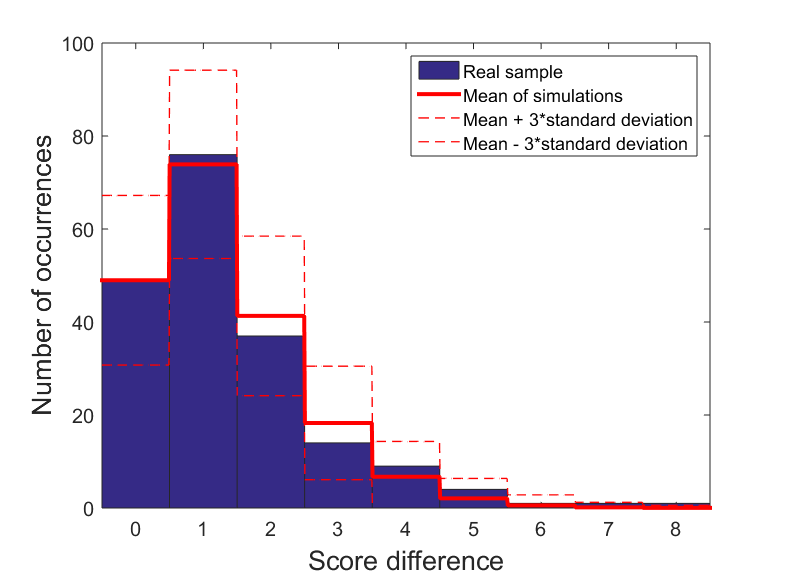}
\caption{Score difference: Poisson model vs. actual data}
\label{histogram}
\end{figure}

\textbf{} \\
We now compare a number of statistics from our chosen Poisson model to those in our sample data at three different levels of aggregation:
\begin{itemize}
\item \textbf{Detailed level}: Each exact final score is considered as a separate event. As our data set is composed of 192 games, we use the Poisson model to carry out 15000 simulations of each of the 192 games. We then count the number of occurrences of each event in each simulation run, and average the results over the 15000 simulations and calculate the standard deviations for each event. The results appear in Table \ref{simtable}, to be compared to the actual outcomes in Table \ref{realtable}. The games in which at least one team scores more than four goals are rare, and do not appear in the tables. The Poisson model provides a good approximation to the actual data.
\item \textbf{Compact level}: Here an event is characterized by the difference in the scores, so that games finishing 3-1 or 4-2 would be similarly categorized as "2-goal difference" events. Figure \ref{histogram} shows the results of our simulations (in red), compared to the actual data (in blue). As above, the Poisson model performs well in reproducing the actual scores.
\item \textbf{Draw frequency}: This is the extreme case where we group all non-zero goal difference outcomes into a single event, with draws being the complement (a goal difference of zero). This allows us to see if the ratio of draws to total number of games in our model matches the sample data. The frequency of draws in the Poisson model is $0.2133$ with a standard  deviation of $0.0297$, while the sample frequency is $0.2552$. As the data only includes $192$ games, we cannot say whether the fit could be improved (by reducing the $a$ parameter, using diagonal-inflated models as per Karlis and Ntzoufras \cite{KN0}, for example) or if our data is not truly representative of the underlying generating process.   
\end{itemize}

Overall, the previous results suggest that the Poisson model fits the score-generating process well.

\begin{table}
\begin{center}
\caption{Number of exact scores in our World Cup data sample of 192 games (goals scored by the team with higher Elo in rows, by the team with lower Elo in columns).}\label{realtable}
\begin{tabular}{|c|ccccc|}
\hline
 \shortstack{Final score} &\textbf{0} & \textbf{1} & \textbf{2} & \textbf{3} & \textbf{4} \\
\hline
\textbf{0} & $14$ & $13$ & $4$ & $2$ & $0$\\
\textbf{1} & $31$ & $23$ & $8$ & $1$ & $0$ \\
\textbf{2} & $18$ & $21$ & $11$ & $2$ & $1$ \\
\textbf{3} & $9$ & $12$ & $1$ & $1$ & $0$ \\
\textbf{4} & $8$& $1$ & $1$ & $0$ & $0$   \\
\hline
\end{tabular}

\caption{Average number of occurrences of exact scores $\pm $ standard deviation based on 15000 Monte Carlo simulations (goals scored by the team with higer Elo in rows, by the team with lower Elo in columns).}\label{simtable}
\begin{tabular}{|c|ccccc|}
\hline
 Final score &\textbf{0} & \textbf{1} & \textbf{2} & \textbf{3} & \textbf{4} \\
\hline
\textbf{0} & $15.6 \pm 3.7$ & $11.0 \pm 3.2$ & $4.6 \pm 2.1$ & $1.4 \pm 1.2$ & $0.3\pm 0.6$\\
\textbf{1} & $27.9 \pm 4.9$ & $18.5 \pm 4.1$ & $7.4\pm 2.7$ & $2.2 \pm 1.5$ & $0.5 \pm 0.7$ \\
\textbf{2} & $26.1 \pm 4.7$ & $15.9 \pm 3.9$ & $6.0 \pm 2.4$ & $1.7 \pm 1.3$ & $0.4 \pm 0.6$ \\
\textbf{3} & $16.6 \pm 3.9$ & $9.2 \pm 3.0$ & $3.4 \pm 1.8$ & $0.9 \pm 1.0$ & $0.2 \pm 0.4$ \\
\textbf{4} & $8.1 \pm 2.8$ & $4.2 \pm 2.0$ & $1.4 \pm 1.2$ & $0.4 \pm 0.6$ & $0.1 \pm 0.3$   \\
\hline
\end{tabular}
\end{center}
\end{table}

%

\newpage

\section{The group-stage match classification method}
\textbf{} \\
We now assess the attractiveness of the group format using the previous model to simulate all the rounds of games in a group, except for the last one. For example, when considering the current format of the World Cup with groups of four countries, we simulate the first two rounds of games (for a total of four games). Then, based on the points system and given that goal difference is the tie-breaker, we calculate the ranking in the group. In case two teams have the same number of points and the same goal difference then they share the same ranking in the group (no other tie-breaker is included in the model).\\

During the last round of games, the qualification of team $i$ will likely not only depend on the outcome of its own game against team $j$ but also on that of the other game, between teams $k$ and $l$. As the last-round games are played simultaneously, we assume that team $i$ does not know the outcome of the other game when playing against $j$. To be precise, from the point of view of team $i$, the $k$ vs $l$ game can end with any (integer) goal difference belonging to the interval $[-5,5]$. We decide to stop at 5 goals difference because modern football games ending with higher goal differences are extremely rare\footnote{Sally and Anderson \cite{Sally} point out that less than 1\% of games in the English Premier League end with a goal difference higher than 5.} (making it likely for team $i$ to disconsider such outcome), and, above all, gives the same numerical results in our simulations as when higher goal differences are taken into account. For each outcome of the $i$ vs $j$ game, we check under which scenarios team $i$ qualifies for the next phase (i.e. ends up in the top two group teams). Occupying a ``clean" first or second position (more points or same number of points but better goal difference than the team in third position) is better than sharing the position with the third-ranked team (additional tie-breakers which are not included in our model may end up disqualifying team $i$). Based on this analysis, team $i$ will choose the lowest-effort outcome that maximizes its chances of qualification: a 5 goals difference win (winning by 5 goals difference is better than winning by 4 goals difference in at least one scenario), ... , a 1 goal difference win (winning by 1 goal difference is no different than winning by more goals difference in all scenarios but better than a draw in at least one scenario), a draw (winning=drawing in all scenarios but is better than losing in at least one scenario), losing by 1 goal difference (same chances of qualification as winning or drawing but is better than losing by 2 goals difference in at least one scenario), ..., and 5 goals difference loss (the team has exactly the same chances of qualifying no matter his performance in the last game). Note that the last situation refers to the case where team $i$ is indifferent since it is already qualified or cannot qualify regardless the result of game $k$ vs. $l$.

For example, suppose that any scenario other than $k$ beating $l$ automatically qualifies $i$. If team $k$ wins, a draw between $i$ and $j$ will qualify $i$, while a loss for $i$ will not lead to qualification (if $i$ wins, it will obviously progress as wins gain more points than draws). In this case, team $i$ will play for a draw: even though team $i$ may still qualify if it loses against $j$, a draw will increase its chances of qualification (if $k$ wins, tying is better for team $i$ than losing). A victory will also qualify $i$, however it does not improve the probability of qualification as compared to a draw. In this same example, if  in a given scenario, a draw leads to a shared second place while a 1 goal difference win leads to a "clean" second place, or even ``clean" first place, then team $i$ will play to win by a 1 goal difference.

Note that, in their strategic choices, teams do not distinguish between first and second place in the group, in the sense that teams only care about qualification to the next round. In the actual World Cup, teams do not always want to finish first in their groups. There have been many occasions where teams seem to have intentionally lost in order to finish second in their group and play against weaker opponents in the knock-out stage. This may well have occurred in the 2018 World Cup in the group with England and Belgium, in which the winner, Belgium, faced more difficult opponents (Brazil and France) in the knock-out stage.   \\

\begin{figure}
\begin{center}
\includegraphics[scale=1]{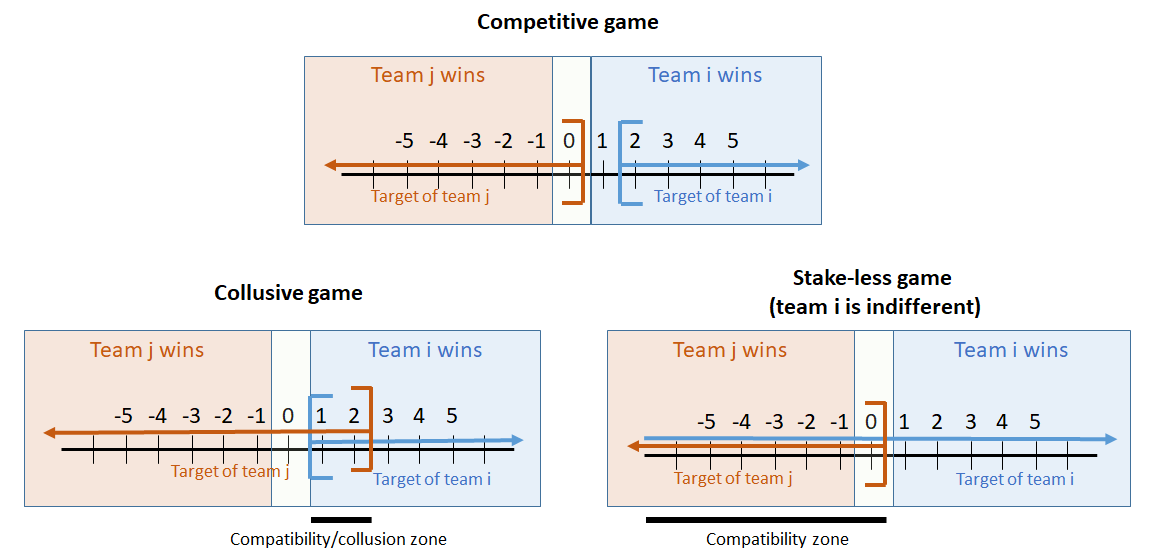}
\caption{Classification of last-round games: competitive, collusive, stake-less (numbers in black represent the final goal difference result of the game in favor of team $i$)}\label{classification_games}
\end{center}
\end{figure}

After having determined what team $i$ would prefer, we carry out an analogous analysis for its opponent $j$, yielding the following classification for the game $i$ vs $j$ as per Figure \ref{classification_games}:
\begin{itemize}
\item \textbf{Competitive games}: Neither team is indifferent here, and their targets are incompatible: if one team reaches its target, the other will not reach theirs. To be clear, both teams may qualify to the second round but there is at least one outcome in the parallel game ($k$ vs $l$) which ``threatens" $i$ and $j$ if their respective targets are not reached. In the example presented in Figure \ref{classification_games}, team $i$ wants to win by a 2 goals difference while team $j$ is looking for a draw. The two teams will thus do their best in this competitive game and the tournament organizer's aim is to ensure that this type of game occurs as often as possible.

\item \textbf{Collusive games}: The targets of both teams are compatible and neither is indifferent: there is a non-empty subset of final scores (in terms of goal difference) which will put the two teams in the best position to qualify (potentially qualifying them at the expense of the other teams of the group). The example presented in Figure \ref{classification_games} is the \textit{Scandal of Gijon}: Germany (team $i$) is looking for a 1 goal difference win while Austria (team $j$) is qualified if it loses by, at most, a 2 goals difference. The ``compatibility/collusion zone" is the subset which contains the 1 goal difference and 2 goals difference victories for Germany (Germany won the game by a 1 goal difference). These types of situations can lead to collusion and should be avoided at any cost.   

\item \textbf{Stake-less games}: At least one of the teams is completely indifferent between winning, drawing or even losing by 5 goals difference. In these games, the indifferent team has, in general, nothing to gain and may field second-team players. This is unfair for the other teams, $k$ and $l$, as they played against a stronger opponent $i$ in the previous rounds. In addition, a team that is already qualified may take into account other factors such as the opponents it will face in the next stage of the competition. Thus, depending on the results in other groups and the scheduling of the next stage, winning the game may not be in its best interest. It can be noticed that these games also present a non-empty compatibility zone (as for collusive games). However, in a collusive set up, both teams are at risk of being disqualified and are expected, ex ante, to play competitively. If the final result is in the compatibility zone, suspicions will immediately be raised even if the two teams did truthfully play competitively. In a stake-less game, spectators and other teams know, ex ante, that the indifferent team will not play competitively (if already qualified, it is his right to put its major players at rest and, if already disqualified, the team can also be encouraged to field young players so they gain experience). Consequently, we strongly believe that stake-less games are less harmful than collusive games but should also be avoided.

\end{itemize}

\section{Assessing the current World Cup format} \label{current_format}
\textbf{} \\
We use the above to assess the last round of the current World Cup format, with groups of four teams of which the top two qualify for the next round. We test both different point-attribution systems and changes to the scheduling of the last round of games:
\begin{itemize}
\item \textbf{Setting 1}: pot $A$ vs. pot $D$ and pot $B$ vs. pot $C$\\
\item \textbf{Setting 2}: pot $A$ vs. pot $C$ and pot $B$ vs. pot $D$\\
\item \textbf{Setting 3}: pot $A$ vs. pot $B$ and pot $C$ vs. pot $D$
\end{itemize}
\vspace{0.5cm}
We carry out 15000 simulations for each setting and point-attribution system: the results appear in Table \ref{MC_table}. Our conclusions are as follows:
\begin{itemize}
\item The points system has no impact on the quality of games (systems with 4-points for a win produce no visible changes in the results)\footnote{It is worth noticing that our model has been calibrated using 3-points for a win, 1-point for a draw and 0-points for a loss data.};
\item Collusive games are relatively rare, but stake-less games are not; and
\item The setting has a considerable impact on the quality of games, with Setting 1 being the best and Setting 3 the worst.

\end{itemize}
The last of the above results is intuitive: in Setting 3, $A$ and $B$ have already played against the weakest teams in the previous rounds. Before the last game, they are likely to have a good number of points, while teams $C$ and $D$ have few or no points. The last round matches the best two teams (who are already or almost qualified) against the weakest two teams (who are already or almost eliminated): the outcome of both games has very little impact on the final group ranking. Furthermore, we have noticed that the only collusion situation which takes place is one in which opposing teams decide to draw. The pressure exerted by the other game which is played in parallel makes a collusion based on goal differences such as the scandal of Gijon very risky (accepting to lose/win but in a given range of goal difference). The introduction of simultaneous matches in the current format of the World Cup has been effective in reducing collusion opportunities.\\

Using our historical data set, we check the proportion of each type (competitive, collusive, stake-less) for the last games of the group round given the scheduling of the group (Setting 1, 2 or 3). The new format has been applied to six World Cups with 8 groups each. The game schedule is drawn randomly, so that each setting is equally likely. Table \ref{data_freq} shows the frequency of each setting in the previous World Cups. We use our classification method to calculate the frequencies of each game type as a function of the setting. Table \ref{data_type} shows the results: Setting 3 produces the least-exciting last round of games, in line with our predictions. Nevertheless, the sample size is only small to check whether the sample estimates fit our model predictions (only 48 last rounds of groups have been played since 1998).

\begin{figure}
\begin{center}
\includegraphics[scale=0.5]{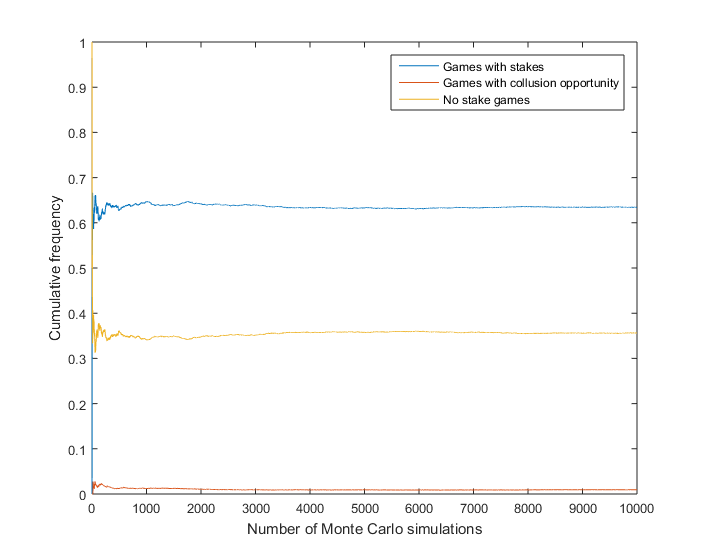}
\caption{The cumulative frequencies in Monte Carlo simulations for a win giving 3 points and a draw 1 point (Setting 1).}\label{MC}
\end{center}
\end{figure}

\begin{table}
\begin{center}
\begin{tabular}{|c|ccc|ccc|}
\hline
\textbf{Setting type :} & 1&2&3 & 1&2&3 \\
\hline
\textbf{Points for :} & \multicolumn{3}{|c|}{\textbf{Win} = 2 }& \multicolumn{3}{|c|}{\textbf{Win} = 3}  \\
\hline
\multirow{3}{*}{\textbf{Draw = 1}} & \textcolor{green_new}{$63.05\%$} & \textcolor{green_new}{$59.50\%$} & \textcolor{green_new}{$42.95\%$}& \textcolor{green_new}{$63.14\%$} & \textcolor{green_new}{$59.49\%$} & \textcolor{green_new}{$42.69\%$}\\
& \textcolor{red_new}{$1.02\%$} & \textcolor{red_new}{$1.15\%$} & \textcolor{red_new}{$1.68\%$} & \textcolor{red_new}{$0.94\%$} & \textcolor{red_new}{$1.32\%$} & \textcolor{red_new}{$1.76\%$}\\
& $35.93\%$ & $39.35\%$& $55.37\%$ & $35.92\%$ & $39.19\%$& $55.55\%$\\
\hline
\multirow{3}{*}{\textbf{Draw = 2}} & \textcolor{green_new}{-} & \textcolor{green_new}{-} & \textcolor{green_new}{-}& \textcolor{green_new}{$63.13\%$} & \textcolor{green_new}{$59.76\%$} & \textcolor{green_new}{$43.54\%$}\\
& \textcolor{red_new}{-} & \textcolor{red_new}{-} & \textcolor{red_new}{-} & \textcolor{red_new}{$0.88\%$} & \textcolor{red_new}{$1.30\%$} & \textcolor{red_new}{$1.60\%$}\\
& - & - & - & $35.99\%$ & $38.94\%$& $54.87\%$\\
\hline
\end{tabular}
\caption{Results of the Monte Carlo simulations with 15000 iterations per run (\textcolor{green_new}{competitive}, \textcolor{red_new}{collusive} and stakeless games).}\label{MC_table}
\end{center}
\end{table}

\begin{table}
\begin{center}
\begin{tabular}{|c|c|c|c|}
\hline
\textbf{Setting} & \textbf{1} & \textbf{2} & \textbf{3}\\
\hline
\textbf{Occurrences} & 15 & 19 & 14\\
\textbf{Frequencies} & $31.25\%$ & $39.58\%$ & $29.17\%$\\
\hline
\end{tabular}
\caption{Number of occurrences and frequencies of the different group settings in our sample data.}\label{data_freq}
\end{center}
\end{table}

\begin{table}
\begin{center}
\begin{tabular}{|c|c|c|c|}
\hline
\textbf{Type of game} & \textbf{Competitive} & \textbf{Stake-less} & \textbf{Collusion opportunity}\\
\hline
\textbf{Setting 1} & $50.00\%$ & $50.00\%$ & $0.00\%$\\
\textbf{Setting 2} & $57.89\%$ & $36.84\%$& $5.26\%$ \\
\textbf{Setting 3} & $32.14\%$ & $67.86\%$ & $0.00\%$\\
\hline
\end{tabular}
\caption{Frequencies of types of games as a function of the group setting.}\label{data_type}
\end{center}
\end{table}

\newpage
\section{The 2026 World Cup}
\textbf{} \\
In 2026, FIFA plans to have 48 qualified teams, distributed into 16 groups of 3 teams. The first two teams in each group (2/3 of teams) then qualify to the knockout stage. The transition from 1/2 to 2/3 teams qualifying has already been tested in the second biggest soccer competition: the UEFA Euro. Between 1996 and 2012, the proportion of teams qualifying for the knockout stage was 1/2 in a 16-team tournament (4 groups of 4 teams), growing to 2/3 for the 2016 UEFA Euro (24 teams divided into 6 groups of 4, first 2 teams per group + 4 best third-ranked teams among all groups qualify). In addition to the change in this ratio, the new FIFA World Cup format will introduce ``passive teams" in the last round of games: in groups of three, one team will have to stand on the side and wait for the result of the last game to decide its fate. Let us notice that, for the new UEFA format, the third ranked teams in groups that have played all their matches also experiment a form of passivity since they have to wait for the outcomes of groups that still have not played their last round of games in order to know if they will qualify. We believe that these changes have led to a decrease in the number of competitive games and would, at least partially, explain the decrease in the number of goals scored in the UEFA Euro Cup (from an average of over 2.5 goals per match in the group phases between 1996 and 2012 to a figure of 1.92 goals per match during the 2016 group phase). This natural experiment then suggests that tournaments with a higher proportion of teams qualifying and/or with passive teams lead to less attractive games and more collusion opportunities. In addition to that, the group-match schedule (the order in which the teams play against each other) may potentially have a critical impact on the quality of the last round games as well.\\

In this last section, taking into consideration FIFA's intention to increase the number of participating teams, we evaluate the new 2026 World-Cup format as well as an  ``augmented UEFA-style" format. Two other alternatives have been analyzed and can be found in the appendices. We should point out that, in addition to having an impact on the ``quality" of games in the FIFA World Cup itself, the new formats have an impact on the ``quality" of games in the continental qualifying tournaments which have to be adapted in order to accomodate the higher number of qualified teams. Such ``secondary" effects of the new World Cup format are not tackled in this article and can set the path for further research.

\subsection{First option: 16 groups of 3}
\textbf{} \\
Some FIFA officials are currently proposing a ``48-team, groups of 3" format, in which the best two teams in each group qualify for the next knock-out stage.
As the groups contain an odd number of teams, one team per group will not play in the last round of games. There are therefore only three possibilities in the last round:
\begin{itemize}
\item \textbf{Setting 1}: The weakest team is the passive team in the last round
\item \textbf{Setting 2}: The middle team is the passive team in the last round
\item \textbf{Setting 3}: The strongest team is the passive team in the last round
\end{itemize}

We carry out 15000 Monte-Carlo simulations to assess the quality of the last round and the results are found in Table \ref{MC_tablegroupsof3}.\\

Two main remarks can be formulated concerning this format\footnote{Similarly to the groups of 4 format of the World Cup, Monte Carlo simulations are performed by randomly (uniformly) drawing Elo indices from 3 consecutive intervals of similar size covering [a,b].} (same conclusions as in Guyon \cite{Guyon2}). First of all, the existence of a passive team re-introduces the \textit{Gijon scandal}-type of collusion opportunities. If the passive team has not accumulated enough points during the first two games of the group, there is a high chance that a win-loss outcome with low goal difference qualifies both the last teams playing. Consequently, the proportion of collusive games is greater in this format than in the current World Cup format (which has no passive team). In the 4-team per group format, there is ``pressure" from the unknown result in the other last-round game which is played simultaneously. This no longer applies in the 3-team per group format.\\
The second point (probably the most consequential) is that the game scheduling has a critical impact on the quality of games in the last round: it is key that the passive team in the last round be the strongest team. In our model, when the strongest team plays the first two rounds of games, there is a $89\%$ chance of a last game in which both teams will give their best. This probability drops to around $68\%$ when the pot $B$ team is the passive one and to $16\%$ when the weakest team is passive ! Indeed, in our model, if the passive team happen to win both its first two games; or draw both his first two games, or win and draw the other or win and lose the other with a total goal difference of zero, then the two non-passive teams will have to play competitively in the last round game. Such scenarios are very likely to take place when the passive team is the pot $A$ team (top pot).\\
\textbf{In case this World Cup format would be carried forward, in order to preserve the fairness and beauty of the game, FIFA should improve its group randomization draw by implementing a predefined schedule in which the pot $A$ team will be the passive team in the last round}.

\begin{table}
\begin{center}
\caption{16 groups of 3 (with 32 qualified teams): Monte Carlo simulations with 15000 iterations.}\label{MC_tablegroupsof3}
\begin{tabular}{|c|ccc|}
\hline
\textbf{Setting type} & \textbf{1} & \textbf{2} & \textbf{3}\\
\hline
\textbf{Interesting games} & $16.07\%$ & $68.40\%$ & $89.09\%$  \\
\textbf{Stakeless games} & $78.62\%$ & $23.45\%$ & $8.17\%$ \\
\textbf{Collusion games} & $5.31\%$ & $8.15\%$ & $2.74\%$\\
\hline
\end{tabular}
\end{center}
\end{table}

\subsection{Second option: 12 groups of 4, 32 qualified to the next stage}
\textbf{}\\
In the spirit of the new format of the UEFA Euro competition, this format consists of 12 groups of 4 teams each. The first 2 teams of each group qualify in addition to the best 8 third-ranked teams among all groups, amounting to a total of 32 qualified teams. \\
First of all, the qualification of the best third-ranked teams creates a considerable informational asymmetry between teams of different groups. For obvious logistic reasons, only a given number of games can be played simultaneously. In the current format of the Euro competition, only the last 2 games of each group are played in parallel. Consequently, the third-ranked team in the first group to play has very few information concerning the number of points and goal difference needed to qualify among the best 8 third-ranked teams. This is not at all the case of the third ranked team of the last group to play since it perfectly knows the number of points and goal difference of the other 11 third-ranked teams, allowing this team to clearly define its target and guarantee its qualification with certainty. In other words, the ``virtual" group of thirds has teams who play sequentially and the first eleven to play become passive teams when the last one plays. This disadvantage for teams playing first is created before any game is played (and even before the final draw since the pot $A$ team of the first group is usually the host of the competition, giving him an actual disadvantage).\\
Since we focus on improving the competitiveness and reducing collusion opportunities in the last round of games, we decide to assess the last round of games for the last group to play. The proportion of non-competitive games will reach its maximum in this last group to play since its third ranked team knows exactly his target to qualify (the results of the previous groups being known to the teams of the last group). In our model, teams look for a ``sure" qualification and do not distinguish between qualifying in 1st, 2nd or 3rd position (among the 8 best thirds). The results of the simulations can be found in table \ref{second_option_uefa}.\\

Compared to the current format of the World Cup, the number of competitive games decreases significantly while collusion opportunities increase. This confirms our intuition that introducing passive teams (namely the 11 teams in the ``virtual" group of thirds) creates collusion opportunities and should be avoided or reduced by increasing the number of parallel games in the last-round of games\footnote{Having 4 games played in parallel could offer a good compromise between having competitive games, not having overwhelming logistic issues, and maximizing viewership (spectators will have to chosen one out of the four simultaneous games).} The other less intuitive result is that the order of optimal game schedules is inversed compared to what we found in Section \ref{current_format}: Setting 3 maximizes the number of competitive games while Setting 1 minimizes it and leads to collusive games $10\%$ of the time. One explanation for this result is that in Setting 1, the pot $B$ team which is likely to be in the 2nd or 3rd position after playing against the strongest and weakest teams of the group, will confront the pot $C$ team in the last round. Pot $C$ team will also likely be in the second or third position since it has played against the strongest and weakest team in its first two games. In this game, a draw or a mild win for the team in the third position may end up qualifying both teams (one qualified among the best 2 teams of the group, the other among the best 8 third-ranked teams). On the other hand, Setting 3 witnesses a confrontation between the two strongest teams and the two weakest ones. It is highly likely that the two weakest ones (pot $C$ vs pot D) will be in 3rd and 4th positions so both teams will  give their best to win the game and qualify among the 8 best third-ranked teams. \textbf{The conclusion is that the ``UEFA Euro"-format leads to more collusion opportunities than the current FIFA World Cup format and, that the last round should witness the confrontation between the strongest two teams and the weakest two in order to minimize collusion opportunities (Setting 3).}

\begin{table}
\begin{center}
\begin{tabular}{|c|c|c|c|}
\hline
\textbf{Type of game} & \textbf{Competitive} & \textbf{Stake-less} & \textbf{Collusion opportunity}\\
\hline
\textbf{Setting 1} & $34.14\%$ & $56.36\%$ & $9.50\%$\\
\textbf{Setting 2} & $37.60\%$ & $55.63\%$& $6.77\%$ \\
\textbf{Setting 3} & $41.71\%$ & $55.47\%$ & $2.82\%$\\
\hline
\end{tabular}
\caption{12 groups of 4 with 32 qualified: results of Monte Carlo simulations with 15000 iterations for the last round of games in the last group to play (settings are the same as for the current World Cup format).}\label{second_option_uefa}
\end{center}
\end{table}

\section{Conclusion}
\textbf{} \\
This article has presented an assessment method of the competitiveness of the last round of games in the FIFA World Cup group stages. We find that in order to reduce the occurrence of match-fixing opportunities, the tournament structure should be optimized so that most of the matches are ``competitive" as per our classification. Applying this new method, we notice that the scheduling of games, in particular the choice of teams playing each other in the last round, is crucial for obtaining exciting and fair last-round games. Furthermore, our results underline that the introduction of passive teams (teams which do not play during the last round of games) increases significantly collusion opportunities and should be avoided, to the extent possible, by scheduling simultaneous games during the last round of the group stage\footnote{In case the number of teams in the group is odd or due to logistical reasons, it may be impossible not to have any passive team.}. The optimal game schedule depends on the format of the World Cup, but our clear recommendation is that FIFA should drop its current schedule-randomization process in the draw for the group matches. In the current World Cup format, we recommend that the last group games should be pot-$A$ teams against pot-$D$ teams, and pot-$B$ teams against pot-$C$ teams. Scheduling these games in advance has no negative impact on any aspect of the competition (including logistics), but reduces the risk of match-fixing and increases the attractiveness and competitiveness of the last round. In the forthcoming 48 teams in groups of 3 format, the ``pot-$A$'' team should be the passive team in the last round. As a path for future research work, it could be interesting to analyze the impact of the half-time break during the last round of games when teams get to know the ``ongoing'' score of other games. Furthermore, psychological effects can be taken into account by including the results of a team's previous matches in the score predicting model (for instance, discouragement after a loss...). Finally, to the best of our knowledge, very few research related to the World Cup continental qualifiers has been done. As FIFA is working towards making the World Cup more exciting and fair, it should not neglect the impact of the new formats of the World Cup on the qualifiers.

\appendix
\section{8 groups of 5 with two qualified per group}
\textbf{} \\
A second alternative is to consider a World Cup with 40 teams divided into 8 groups of 5 teams. The first two teams in each group qualify for the knock-out stage. In this new format, each team will have to play one more game in the group stage, and the last round of games will also include a ``passive" team due to the odd number of teams per group.
This passive team will have played all its games previously and will have to watch the other four teams play during the last round.

As in the previous cases, we carry out a Monte Carlo simulation to determine which combination of games should be played in the last round. The results appear in Table \ref{MC_table_groupsof8}, where the attribute ``game quality" is that defined previously, ``passive team" is the team that has already played four games and ``setting" corresponds to the matching combination of the four teams left.
The results are the following:
\begin{itemize}
\item As for the current World Cup format, Setting 3 produces worse results than settings 1 and 2. This is the one in which the strongest teams left play against each other, and the weakest two play each other.
\item The choice of the passive team affects the quality of results via two mechanisms:
\begin{enumerate}
\item The stronger the passive team, the more chances it has to win its games and be qualified for the next round. There will consequently only be one spot left for the other four teams playing in the last round. This increases the chances of there being some teams that cannot qualify among the four left, leading to uninteresting last-round games. \textbf{Based on this mechanism, the weaker the passive team the better the quality of last-round games.}
\item The second effect is related to the position of the four teams left relative to the passive team. If the passive team is the weakest ``pot E" (or the strongest ``pot A"), then it is likely that the four teams left had positive (negative) results against the passive team. What matters here is not whether they won or lost, but that these four teams had the same outcome against the passive team. They thus start the last-round games with the same head start, leading to competitive last-round games. On the contrary, if the passive team is from pot $C$, the teams coming from pots $A$ and $B$ were likely to have won against it, while the teams from pots $D$ and $E$ were likely to have lost. As such, the last round starts with a considerable gap between the teams in terms of points, so that some teams may be already qualified (or not be able to qualify). This phenomenon stands out clearly in Setting 3, where the pots $A$ and $B$ teams have played against the weakest teams and play each other in the last round. It is likely that they have both already qualified before the last round, making this game stakeless. \textbf{Consequently, the more extreme is the passive team (pots $A$ or E), the better is the last round of games.}    
\end{enumerate}
\end{itemize}

Based on the previous results, we conclude that Setting 3 should be avoided (Setting 1 is the optimal one), while the pot-E team should be the passive team in the last round. For a similar number of starting teams, 40 teams versus 48 teams in the second option ("UEFA"-format), the optimized version of this format leads to more competitive games, less collusion opportunities, and no ex-ante disadvantage across groups.

\begin{center}
\begin{table}
\caption{8 groups of 5 with 16 qualified teams (2 per group):  Monte Carlo simulation with 15000 iterations. (Setting corresponds to the matching combination of the four teams left)}\label{MC_table_groupsof8}
\begin{tabular}{|c|ccc|ccc|ccc|}
\hline
\textbf{Setting:} & \multicolumn{3}{|c|}{ \textbf{1} }& \multicolumn{3}{|c|}{\textbf{2}} & \multicolumn{3}{|c|}{\textbf{3}} \\
\hline
\textbf{Game quality:} & \textbf{Comp} & \textbf{Col} & \textbf{Stkless} &  \textbf{Comp} & \textbf{Col} & \textbf{Stkless} &  \textbf{Comp} & \textbf{Col} & \textbf{Stkless} \\
\hline
\textbf{Passive team} & \multicolumn{9}{|c|}{}\\
\hline
\textbf{Pot A} & $35.58\%$ & $0.41\%$ & $64.01\%$ & $35.11\%$ & $0.76\%$ & $64.14\%$ & $34.17\%$ & $1.88\%$ & $63.95\%$\\
\hline
\textbf{Pot B} & $33.15\%$ & $0.36\%$ & $66.48\%$ & $31.27\%$ & $1.26\%$ & $67.47\%$ & $28.63\%$ & $2.46\%$ & $68.91\%$\\
\hline
\textbf{Pot C} & $34.05\%$ & $0.63\%$ & $65.31\%$ & $32.37\%$ & $1.02\%$ & $66.61\%$ & $22.77\%$ & $3.40\%$ & $73.83\%$\\
\hline
\textbf{Pot D} & $38.93\%$ & $1.26\%$ & $59.81\%$ & $34.72\%$ & $1.94\%$ & $63.34\%$ & $27.27\%$ & $2.80\%$ & $69.93\%$\\
\hline
\textbf{Pot E} & $57.35\%$ & $1.64\%$ & $41.01\%$ & $54.91\%$ & $1.84\%$ & $43.25\%$ & $47.17\%$ & $2.67\%$ & $50.16\%$\\
\hline
\end{tabular}
\end{table}
\end{center}

\section{12 groups of 5, 32 qualified teams to the next stage}
\textbf{} \\
In this variant of the previous format, the first two teams of each group as well as the 8 best third-ranked teams across all groups qualify. As for the "UEFA-format", this format leads to ex-ante disadvantage between groups and table \ref{fourth option} shows the results of our simulations for the last round of games in the last group to play.\\
As we can expect, the introduction of passive teams (due to the qualification of the 8 best thirds) causes a deterioration in the quality of games compared to the previous alternative (for all settings). Furthermore, similarly to what happened when we add the qualification of the best thirds, the order of optimal settings is modified. When the passive team is from pot E, Setting 1 becomes the least optimal while Setting 3 becomes the most optimal.\\
Such an alternative format should be discarded but confirms that the introduction of passive teams (qualifying the best 8 third-ranked teams) increases the chances of collusive games.

\begin{center}
\begin{table}
\caption{12 groups of 5 with 32 qualified teams: results of the Monte Carlo simulations with 15000 iterations for the last round of games in the last group to play. (Setting corresponds to the matching combination of the four teams left)}\label{fourth option}
\begin{tabular}{|c|ccc|ccc|ccc|}
\hline
\textbf{Setting:} & \multicolumn{3}{|c|}{ \textbf{1} }& \multicolumn{3}{|c|}{\textbf{2}} & \multicolumn{3}{|c|}{\textbf{3}} \\
\hline
\textbf{Game quality:} & \textbf{Comp} & \textbf{Col} & \textbf{Stkless} &  \textbf{Comp} & \textbf{Col} & \textbf{Stkless} &  \textbf{Comp} & \textbf{Col} & \textbf{Stkless} \\
\hline
\textbf{Passive team} & \multicolumn{9}{|c|}{}\\
\hline
\textbf{Pot A} & $34.01\%$ & $2.15\%$ & $63.85\%$ & $29.09\%$ & $2.66\%$ & $68.25\%$ & $18.79\%$ & $3.61\%$ & $77.60\%$\\
\hline
\textbf{Pot B} & $31.98\%$ & $2.46\%$ & $65.55\%$ & $24.36\%$ & $2.99\%$ & $72.66\%$ & $15.17\%$ & $3.23\%$ & $81.59\%$\\
\hline
\textbf{Pot C} & $26.09\%$ & $3.51\%$ & $70.40\%$ & $23.18\%$ & $3.42\%$ & $73.40\%$ & $11.36\%$ & $2.79\%$ & $85.86\%$\\
\hline
\textbf{Pot D} & $19.55\%$ & $4.64\%$ & $75.81\%$ & $17.41\%$ & $3.95\%$ & $78.64\%$ & $14.68\%$ & $2.97\%$ & $82.35\%$\\
\hline
\textbf{Pot E} & $31.97\%$ & $6.29\%$ & $61.74\%$ & $32.28\%$ & $5.71\%$ & $62.01\%$ & $32.55\%$ & $4.13\%$ & $63.31\%$\\
\hline
\end{tabular}
\end{table}
\end{center}

\end{document}